\documentclass[%
reprint, 11pt, onecolumn,
 amsmath,amssymb,
 aps,prb,nobalancelastpage
]{revtex4-2}

\usepackage{graphicx}% Include figure files
\usepackage{dcolumn}% Align table columns on decimal point
\usepackage{bm}% bold math
\usepackage{balance}
\usepackage{comment}
\usepackage{amsmath}
\usepackage{xcolor}

\begin{document}
\title{Robust Quantum Learning through Hamiltonian Reservoir Computing}

\author{Youya Xu}
\affiliation{School of Science and Engineering, The Chinese University of Hong Kong (Shenzhen), Longgang, Shenzhen, Guangdong, 518172, P.R. China}

\author{Chengyong Yu}
\affiliation{School of Science and Engineering, The Chinese University of Hong Kong (Shenzhen), Longgang, Shenzhen, Guangdong, 518172, P.R. China}

\author{Sanjib Ghosh}
\email{sanjibghosh@cuhk.edu.cn}
\affiliation{School of Science and Engineering, The Chinese University of Hong Kong (Shenzhen), Longgang, Shenzhen, Guangdong, 518172, P.R. China}
\affiliation{Beijing Academy of Quantum Information Sciences, Beijing, 100193, P. R. China}

\begin{abstract}
Quantum learning provides a versatile paradigm for information processing by exploiting the intrinsic representational capacity of high-dimensional Hilbert spaces. Here, we investigate a Hamiltonian-encoding framework for quantum reservoir computing that simultaneously addresses three key challenges in quantum learning: trainability, hardware efficiency, and information stability. In this framework, input data are directly mapped onto a fixed Hamiltonian and transformed into expressive nonlinear features through quantum dynamical evolution. By employing the reservoir-computing paradigm, the approach naturally circumvents the barren plateau problem in quantum learning landscapes. We validate the framework across two complementary platforms: an analog superconducting array processor and a digital gate-based quantum circuit implementation. Despite their fundamentally different realizations, both platforms exhibit comparable representational power and achieve competitive learning performance, establishing a unified framework for cross-platform quantum learning. While both implementations achieve comparable performance, the analog processor may offer a more hardware-efficient realization by bypassing the temporal overhead of gate-based decomposition and thereby making more effective use of finite coherence times, albeit at the expense of universality. Furthermore, we find that finite dissipation suppresses quantum-scrambling-induced instabilities at long evolution times and can enhance learning performance, revealing a constructive role for environmental coupling in stabilizing quantum learning dynamics. Collectively, these results establish Hamiltonian-encoded reservoir computing as a compact, expressive, and hardware-efficient paradigm for quantum learning on current-generation quantum platforms.
\end{abstract}
\maketitle

\section{Introduction}
Quantum computing promises computational advantages over classical algorithms for certain tasks, including quantum learning, which exploits the exponential dimensionality of Hilbert space~\cite{Schuld2021Machine}. However, realizing practical quantum learning advantages remains challenging and is still under active investigation, as the exponentially large Hilbert space of quantum systems does not by itself guarantee superior representational capacity or robust generalization~ \cite{Cerezo2022Challenges}. In practice, many contemporary quantum learning models are constrained by the requirement for deep circuit architectures or extensive system scaling to achieve competitive performance, often encountering significant bottlenecks such as barren plateaus and high gate overhead~\cite{mcclean2018barren, holmes2022connecting}.

To circumvent these limitations, several approaches have emerged, ranging from quantum feature embeddings~\cite{Rebentrost2014} to variational quantum circuits~\cite{kandala2017hardware}. It has been demonstrated that quantum feature-space methods, such as variational classifiers and kernel estimators, can be realized on near-term superconducting hardware~\cite{havlivcek2019supervised}. Complementing these trainable architectures, a parallel research direction has shifted toward fixed dynamical feature generation. This approach is closely related to the reservoir computing paradigm, wherein a high-dimensional nonlinear dynamical system acts as a fixed reservoir, restricting the training process to the output readout. Classical echo state networks demonstrated that fixed recurrent dynamics can transform input sequences into high-dimensional representations, simplifying learning to a linear task~\cite{jaeger2004harnessing}. 
Recent advances in next-generation reservoir computing have further established that this learning efficiency is centrally driven by nonlinear feature expansion~\cite{Luko2009Reservoir, gauthier2021next, Aadhi2025}.

Quantum reservoir computing extends this paradigm by substituting the classical reservoir with quantum systems, thereby exploiting the vast Hilbert space and intrinsic quantum evolution for feature generation~\cite{fujii2017harnessing,ghosh2019quantum,Mujal2021}. This approach is conceptually aligned with quantum feature mapping, which embeds classical data into high-dimensional quantum states for subsequent classification~\cite{schuld2019quantum, havlivcek2019supervised}. Building upon this underlying framework, recent experimental demonstrations have successfully validated such fixed quantum architectures as viable physical substrates for machine learning tasks~\cite{Suprano2024, senanian2024microwave,hou2026high,paparelle2026experimental,cimini2026large}.

Despite recent advances in quantum learning, several fundamental challenges remain unresolved. First, many existing approaches rely on variational optimization, which is often hindered by barren plateaus and unfavorable training landscapes, limiting scalability and trainability~\cite{cerezo2021cost, anschuetz2022traps}. Second, most quantum learning architectures are closely tied to specific hardware platforms, making it difficult to establish a unified framework for cross-platform implementation, validation, and performance benchmarking across analog and digital quantum processors. Third, quantum information scrambling in many-body systems can induce learning instabilities and substantially degrade performance during extended dynamical evolution~\cite{Xiong2025, Xiong2025Role, Kobayashi2026}. Consequently, it remains an open question whether quantum devices can achieve robust and competitive learning performance while operating within the resource constraints of current hardware.

Here, we investigate a Hamiltonian Encoding Framework (HEF) tailored for reservoir computing, in which input data are directly mapped onto a fixed Hamiltonian and subsequently transformed into nonlinear features through quantum dynamical evolution~\cite{Harrow2009,McCaul2025}. We show that the resulting feature map is highly expressive, exhibiting both a high-dimensional variance structure and a robust local nonlinear response. Using a Hilbert-space dimension corresponding to only a few qubits, the HEF achieves competitive test accuracies on the full MNIST dataset~\cite{Lecun1998}, demonstrating that high-performance quantum learning can be attained with remarkably limited quantum resources. Moreover, because the HEF is rooted in the reservoir-computing paradigm, learning is performed solely through a classical readout layer, naturally avoiding the costly optimization landscapes and barren plateau problems that often limit variational approaches~\cite{cerezo2021cost, anschuetz2022traps}.

To address the challenge of platform-specific quantum learning architectures, we demonstrate that the HEF can be implemented across both analog and digital quantum processors while maintaining comparable performance. The analog realization employs an Analog Superconducting Array Processor (ASAP), for which we perform large-scale numerical simulations using a Hamiltonian tailored to superconducting-qubit architectures. Our systematic investigation shows that the analog processor achieves MNIST test accuracies approaching 98\% using only five to six qubits.

We further demonstrate that the HEF admits an efficient digital realization through shallow quantum circuits whose depth scales linearly with the input dimensionality. The resulting Quantum Circuit Implementation (QCI) consists of an encoding block that embeds the data and an interaction block that generates the required qubit mixing. Using this architecture, we obtain MNIST test accuracies of approximately 97.5\%, establishing near-parity with the analog implementation.

Although the ASAP and QCI exhibit comparable representational performance, the analog processor achieves this performance within substantially shorter evolution times. In contrast, the digital realization incurs additional temporal overhead associated with decomposing continuous many-body dynamics into discrete gate operations. Consequently, the ASAP provides a more hardware-efficient realization of the HEF, maximizing the effective utilization of coherence times in superconducting quantum processors.

Finally, we investigate the impact of environmental coupling on learning performance and address the challenge posed by information scrambling in quantum reservoir computing. We find that performance at short evolution times remains remarkably stable even in the presence of significant dissipation. More surprisingly, at longer evolution times, finite dissipation enhances learning performance by suppressing scrambling-induced instabilities. These results reveal a constructive role for controlled dissipation in stabilizing quantum learning dynamics. Collectively, our findings establish Hamiltonian-encoded reservoir computing as a compact, expressive, and hardware-efficient paradigm for quantum learning on near-term superconducting quantum platforms.

\textit{Organization of the Paper}: We organize this paper to reflect the progression from the theoretical formulation of the HEF to its practical implementation on quantum hardware. Section~\ref{sec:theory1} establishes the theoretical foundations, detailing the mechanisms for data encoding, Hamiltonian evolution, and feature extraction. We then present two distinct implementation architectures: the Analog Superconducting Array Processor (ASAP) in Section~\ref{sec:theory2} and the Quantum Circuit Implementation (QCI) in Section~\ref{sec:theory3}, each including specific methodological details and benchmark results. Section~\ref{sec:comparison} provides a comparative analysis of these two realizations. Section~\ref{sec:opensystem} evaluates the framework's performance under realistic dissipative conditions and discusses the non-trivial performance advantages afforded by open-system dynamics. Finally, Section~\ref{sec:discussion} summarizes the primary findings and explores the broader implications of this work.

\section{Theoretical Framework}\label{sec:theory1}
 
Here we consider a quantum learning framework based on Quantum Reservoir Processing (QRP)—a quantum-mechanical generalization of classical reservoir computing. Within this framework, a fixed, untrained quantum network maps input data into an exponentially large Hilbert space, leveraging the rich many-body Hamiltonian dynamics for feature representation. Although quantum reservoirs provide a large feature space, the identification of a robust data encoding protocol remains an open challenge. While various strategies, including phase, amplitude, and continuous-variable encoding, have been explored in diverse contexts, the identifying an optimal scheme for information injection is still a subject of active research.

\begin{figure*}
    \centering
    \includegraphics[width=\linewidth]{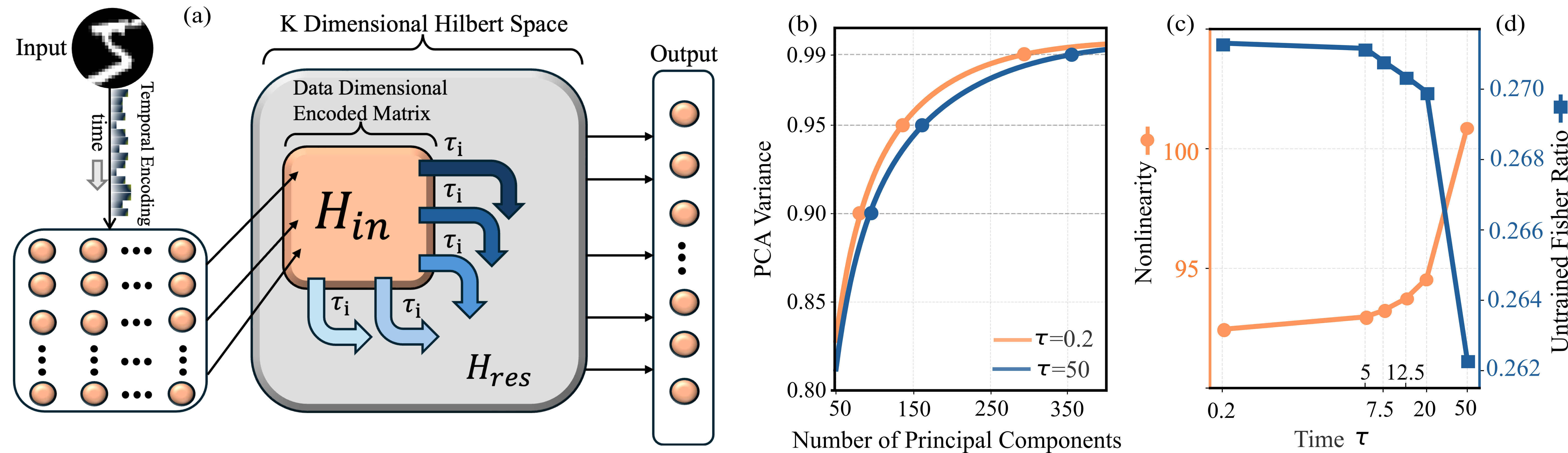}
    \caption{ 
    \textbf{Schematic diagram and expressivity analysis of the Hamiltonian Encoding Framework (HEF)}. 
    \textbf{(a)} Schematic representation of the HEF architecture. Input data is mapped via temporal encoding into a $K$-dimensional Hamiltonian, which undergoes unitary evolution in the Hilbert space across multiple temporal scales (multiplexing). The resulting dynamical features are processed in the output layer to determine the optimized weights $\mathbf{W}_{\mathrm{out}}$ for classification. 
    \textbf{(b)} Dimensionality analysis via PCA cumulative explained variance in $\tau=0.2$ and $\tau=50$. The solid-circles denote the number of principal components required to capture $90\%$, $95\%$, and $99\%$ of the total variance, respectively. 
    \textbf{(c)} Local nonlinearity strength of the HEF feature map, evaluated as the mean second-order response under small input perturbations. 
    \textbf{(d)} Untrained class separability quantified by the Fisher ratio, defined as the ratio of the traces of the between-class scatter matrix to the within-class scatter matrix. While the high local nonlinearity confirms the system's task-agnostic expressivity, this strong state-space folding naturally leads to a relatively small untrained Fisher ratio prior to readout optimization. Energies and timescales are expressed in units of $100\,\text{MHz}$ and $10\,\text{ns}$, respectively.}
    \label{fig:mn1}
\end{figure*}

In this work, we demonstrate that Hamiltonian encoding provides a robust and effective framework for QRP. Fig.~\ref{fig:mn1} presents a schematic overview of the Hamiltonian Encoding Framework (HEF) and characterizes the representational diversity of the generated quantum features. In the HEF, we embed the input data directly into the reservoir Hamiltonian $H$. The subsequent time evolution $e^{-iHt}$ acquires a highly input dependent structure in the Hilbert space and yields rich data representations. Moreover, since the quantum evolution operator is an oscillatory function of the Hamiltonian, the mapping between the input data and the output measurements is strongly nonlinear which enhances the feature space and therefore provides a clear advantage for classification~\cite{schuld2019quantum,havlivcek2019supervised}.

However, the high degree of nonlinearity inherent in the HEF often results in a highly complex training landscape, which can impede learning efficiency. While gradient-based optimization is the standard for navigating loss functions in conventional networks, its application to quantum neural networks is frequently hindered by rugged landscapes and nearly flat regions. These `Barren plateaus'~\cite{McClean2018BarrenPlateaus}, where gradients vanish exponentially, render effective training impractical. Consequently, we investigate whether the synergy between the QRP and HEF can circumvent these challenges to provide a stable and computationally efficient learning platform.

\subsection{Hamiltonian Construction}

In this section, we consider a mathematical model for the HEF in QRP to establish a robust working protocol, before introducing two complementary physical implementations. For mathematical abstraction, we consider that the reservoir is represented by a Hamiltonian matrix $\mathbf{H}_{\mathrm{res}}$ part of which directly contains the input data. The Hamiltonian is represented by a matrix $\mathbf{H}_{\mathrm{res}} \in \mathbb{C}^{K \times K}$, where $K$ denotes the dimension of the Hilbert space, which is constructed for the dynamical feature generation. Keeping the physical set-up and the requirements of feature representation in mind, we defined $\mathbf{H}_{\mathrm{res}}$ as:

\begin{equation}
[\mathbf{H}_{\mathrm{res}}]_{ij} =
\begin{cases}
[\mathbf{H}_{\mathrm{in}}]_{ij}, & 1 \le i,j \le d,\\[4pt]
[\mathbf{H}_{\mathrm{base}}]_{ij}, & \text{otherwise},
\end{cases}
\end{equation}
where $d$ is the input data size, $\mathbf{H}_{\mathrm{in}}$ represents the input-dependent block, and $\mathbf{H}_{\mathrm{base}}$ is a fixed background Hamiltonian. 
We note that QRP typically does not depend on the specific form of the reservoir, but instead relies on the global properties of the dynamics it generates. Consequently, we can expect considerable flexibility in the detailed structure of the Hamiltonian without affecting the final performance.

To establish a general framework, we adopt a formal procedure to construct the input part of the Hamiltonian $\mathbf{H}_{\mathrm{in}}$, which is constructed from classical data through a linear mapping. First, the input image is flattened into a vector $\mathbf{x}\in\mathbb{R}^{d^2}$. The initial encoding is defined as:
\begin{equation}
\label{eq:input_encoding}
\tilde{\mathbf{x}} = \mathbf{W}_{\mathrm{in}} \mathbf{x} + 0.05\,\mathbf{b},
\end{equation}
where $\mathbf{W}_{\mathrm{in}}\in\mathbb{R}^{d^2 \times d^2}$ is a fixed random linear map and $\mathbf{b}\in\mathbb{R}^{d^2}$ is a fixed bias vector. To ensure numerical stability and a consistent energy scale, the encoded vector is normalized: $\mathbf{x}_{\mathrm{in}}= {\tilde{\mathbf{x}}}/{|\tilde{\mathbf{x}}|}$. This normalized vector is then reshaped into a $d\times d$ square matrix: $\mathbf{B}_{\mathrm{in}} = \mathrm{reshape}(\mathbf{x}_{\mathrm{in}}, d, d)$. To satisfy the requirement of Hermiticity for the input block, we apply a symmetrization step:
\begin{equation}
\label{eq:input_block_hermitian}
\mathbf{H}_{\mathrm{in}}=\frac{1}{2}\left(\mathbf{B}_{\mathrm{in}}+\mathbf{B}_{\mathrm{in}}^\dagger\right).
\end{equation}
The input data is integrated directly into the reservoir via $\mathbf{H}_{\mathrm{in}}$, which serves as a local perturbation to the global Hamiltonian landscape.

The reservoir network is characterized by a fixed, untrained background operator $\mathbf{H}_{\mathrm{base}}$, modeled as a random Hermitian matrix:
\begin{equation}
\label{eq:base_hamiltonian}
\mathbf{H}_{\mathrm{base}}=\frac{1}{2}\left(\mathbf{A}+\mathbf{A}^\dagger\right),
\end{equation}
where $\mathbf{A}\in\mathbb{C}^{K\times K}$ is a complex random matrix with elements independently sampled as

\begin{equation}
\label{eq:random_complex_matrix}
A_{ij}=2\left[\left(r_{ij}-\frac{1}{2}\right)+i\left(s_{ij}-\frac{1}{2}\right)\right]; \quad r_{ij}, s_{ij}\sim \mathcal{U}(0,1),
\end{equation}
for $i,j=1,\dots,K$. This construction is equivalent to drawing the real and imaginary components of $A_{ij}$ independently from a uniform distribution on the interval $[-1,1]$. The Hermitization in Eq.~\eqref{eq:base_hamiltonian} ensures that $\mathbf{H}_{\mathrm{base}}$ is a self-adjoint operator, thereby providing a physically consistent unitary evolution.

While $\mathbf{H}_{\mathrm{base}}$ defines the static structural substrate of the reservoir, sample-specific information is introduced via a local modification: the top-left block of the background Hamiltonian is substituted with the input-encoded block $\mathbf{H}_{\mathrm{in}}$. This mechanism ensures that while the internal connectivity of the  reservoir remains constant, each input data point uniquely reconfigures the local energy landscape.

To ensure dynamical stability and preclude divergent behavior, we rescale the spectral radius of the Hamiltonian to a fixed value. This spectral normalization is the quantum analogue of the echo-state property in classical reservoir computing~\cite{jaeger2004harnessing} and serves as a critical control parameter for maintaining stability in quantum reservoir dynamics.
\begin{equation}
\label{eq:spectral_control}
\mathbf{H}_{\mathrm{res}} \leftarrow \mathbf{H}_{\mathrm{res}}   \frac{r_{\mathrm{tr}}}{r_{\mathrm{cr}}}, \quad r_{\mathrm{cr}} = \max(|\lambda_i|)
\end{equation}
where $r_{\mathrm{tr}} = 0.88$ represents the target spectral radius, $\lambda_i$ are the eigenvalues of $\mathbf{H}_{\mathrm{res}}$, and $r_{\mathrm{cr}}$ is the current spectral radius (all the energies including $r_{\mathrm{cr}}$ are expressed in the unit of $100\,$mHz). This normalization provides a practical balance between dynamical richness and numerical stability.

%%%%%%%%
\subsection{Time Evolution}
The time evolution of an initially prepared quantum state of the quantum reservoir is given by the von Neumann equation:
\begin{equation}
\label{eq:vonNeumann}
i\frac{d\boldsymbol{\rho}}{dt}
=[\mathbf{H}_{\mathrm{res}},\boldsymbol{\rho}]
\end{equation}
where $H_\text{res}$ is the Hamiltonian, $\rho$ is the density matrix of the system, and $[.,.]$ denotes the commutator. For evolution, we consider natural units where $\hbar = 1$. Since, reservoir computing does not demand any particular structure in the initial states, we prepare them with statistical ensemble procedure. In this procedure, randomly generated quantum ensemble states $\{|\psi_{i,0}\rangle\} \in \mathbb{C}^{K}$ are mixed through the convex combination:
\begin{equation}
\label{eq:initial_density_matrix}
\boldsymbol{\rho}_0 = \frac{1}{M}\sum_{i=1}^{M} |\psi_{i,0}\rangle \langle \psi_{i,0}|\,,
\end{equation}
where the individual state is normalized. By varying the number of states $M$, we obtain different initial purity. Notably, the initial state $\rho_0$ does not encode the input data, but it enters the framework only through the Hamiltonian $\mathbf{H}_{\mathrm{res}}$.

The von Neumann equation governs the coherent quantum dynamics driven by the Hamiltonian $\mathbf{H}_{\mathrm{res}}$. For an evolution time $\tau$, the density matrix evolves as
\begin{equation}
\boldsymbol{\rho}(\tau) = \mathbf{U}(\tau) \boldsymbol{\rho}_0 \mathbf{U}^\dagger(\tau),
\end{equation}
where $\mathbf{U}(\tau) = \exp(-i \mathbf{H}_{\mathrm{res}} \tau)$ is the unitary evolution operator.
Since quantum evolution is highly nonlinear in the Hamiltonian parameters where the information is encoded, the evolution operator generates a vast number of features at different propagation times $\tau \in \{\tau_1, \tau_2, \ldots, \tau_n\}$. We utilize this temporal embedding to generate a high-dimensional feature space from a single reservoir Hamiltonian $H_\text{res}$. The set of propagation times $\{\tau_j\}$ is chosen to capture both short-time quantum coherence, where phase correlations are largely preserved, and long-time dynamics characterized by strong mixing~\cite{dalessio2016quantum}.

\subsection{Feature Extraction Protocol}
Within this framework, an process is defined as the generation of features through a single quantum evolution. For each discrete evolution cycle $i$, the system originates from an initial state $\rho_0$ and evolves for a duration $\tau_i$, after which we extract physically meaningful observables. The diagonal elements of the resulting density matrix, representing the state populations, are given by:
\begin{equation}
\mathcal{D}_i = \mathrm{diag}[\rho(\tau_i)] \in \mathbb{R}^K
\end{equation}
where the mapping $\mathrm{diag}[\cdot]$ extracts the diagonal entries of the density matrix. Physically, these elements correspond to the populations in the local measurement basis. From an implementation perspective, measurement in the computational basis represents the fundamental lower bound of complexity in quantum computing; it is the uniquely optimal scheme that requires no pre-measurement unitary rotations and exhibits a sample complexity that scales only linearly with the Hilbert space dimension.

Moreover, in the pure-state limit, the magnitudes of the off-diagonal elements (coherence) can be directly inferred from basis measurements via the relation $|\rho_{jk}| = \sqrt{\rho_{jj} \rho_{kk}}$. Consequently, within the pure-state processing domain, basis measurements alone provide the following feature set:
\begin{equation}
\mathbf{f}_\psi (\tau_i) = [ \{\rho_{jj}(\tau_i)\}, \{|\rho_{jk}(\tau_i)|\}]
\end{equation}
for all $j = 1, 2, \dots, K$ and $k > j$. The total number of independent features extracted from a single process is therefore $K(K+1)/2$.

Beyond basis measurements, when Informationally Complete (IC) measurements are accessible, the full set of $K^2$ independent parameters within the density matrix can be extracted. In this regime, we define the feature set by the unique real and imaginary components of $\rho$:
\begin{equation}
\mathcal{R}_i = \mathrm{Re}[\rho(\tau_i)_{jk}],~~~~
\mathcal{I}_i = \mathrm{Im}[\rho(\tau_i)_{jk}]
\end{equation}
where the operators $\mathrm{Re}[\cdot]$ and $\mathrm{Im}[\cdot]$ extract the respective parts for $j=1,\dots, K$ and $k>j$. This set comprises $K$ diagonal elements and $K(K-1)$ independent off-diagonal real-valued features. The complete feature vector $\mathbf{f}_{\rho}$ characterizing the reservoir state is thus constructed as:
\begin{equation}
\label{eq:full_feature_vector}
\mathbf{f}_{\rho}(\tau_i) = \big[ \mathcal{D}_i, \vec{\mathcal{R}}_i, \vec{\mathcal{I}}_i \big]
\end{equation}
where $\mathcal{D}_i$ represents the diagonal populations, and $\vec{\mathcal{R}}_i$ and $\vec{\mathcal{I}}_i$ denote the vectorized form of the independent off-diagonal components. The total dimensionality of the feature space per process is $K^2$.

\subsection{Temporal Concatenation}\label{sec:heffeatures}

Features from $n$ discrete evolution processes are concatenated to form a comprehensive representation of the system state. The dense feature vector, which incorporates the full informationally complete state, is constructed as:
\begin{equation}
\mathbf{x}_{\mathrm{dense}} = \big[ 1, \, \mathbf{x}_{\mathrm{in}}, \, \mathbf{f}_{\rho}(\tau_1), \, \dots, \, \mathbf{f}_{\rho}(\tau_n) \big]
\end{equation}
where $\mathbf{x}_{\mathrm{in}}$ denotes the input data block. This concatenation strategy assembles a high-dimensional feature space spanning multiple evolution timescales, providing complementary dynamical information for subsequent prediction tasks. Including the constant bias term and the input block $\mathbf{x}_{\mathrm{in}} \in \mathbb{R}^{d^2}$, the total dimension of the dense feature space is:
\begin{equation}
\dim(\mathbf{x}_{\mathrm{dense}}) = 1 + d^2 + nK^2
\end{equation}

In the limit of pure-state processing via basis measurements, while the underlying Hamiltonian construction remains invariant, the state representation is modified to utilize the population-based feature set:
\begin{equation}
\label{eq:final_feature_wavefunction}
\mathbf{x}_{\mathrm{basis}} = \left[ 1, \, \mathbf{x}_{\mathrm{in}}, \, \mathbf{f}_{\psi}(\tau_1), \, \dots, \, \mathbf{f}_{\psi}(\tau_n) \right]
\end{equation}
The resulting total feature dimension used for training is:
\begin{equation}
\dim(\mathbf{x}_{\mathrm{basis}}) = 1 + d^2 + n \frac{K(K+1)}{2}
\end{equation}
Consequently, while the dense feature representation $\mathbf{x}_{\mathrm{dense}}$ leverages the full informationally complete state space for maximum expressive power, the basis-measurement approach $\mathbf{x}_{\mathrm{basis}}$ provides a hardware-efficient alternative that retains essential quadratic scaling of the feature space while operating at the fundamental lower bound of measurement complexity. 

\begin{figure*}
    \centering
    \includegraphics[width=0.95\linewidth]{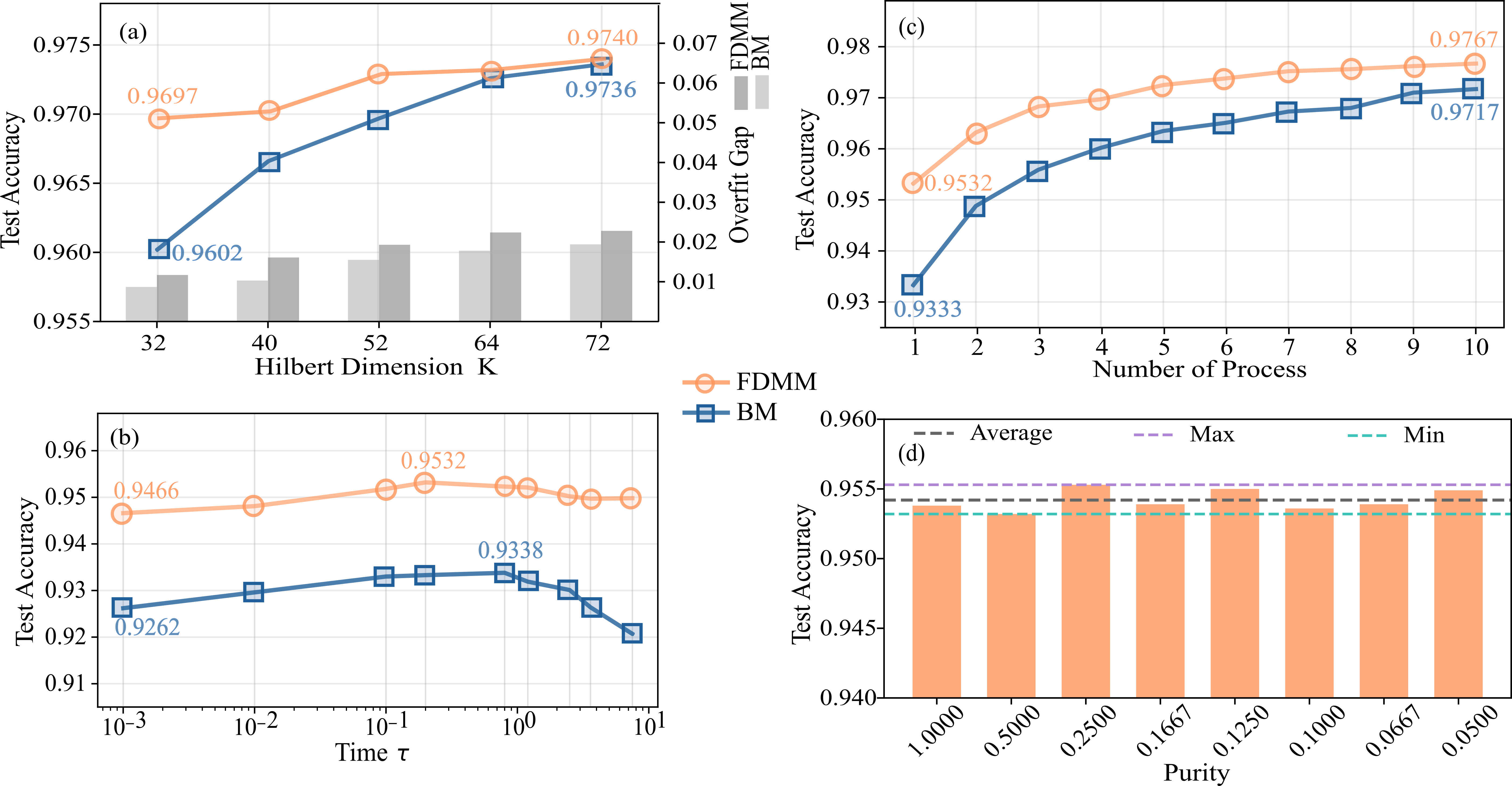}
    \caption{\textbf{Performance scaling and parameter optimization of the HEF framework.} 
\textbf{(a)} MNIST classification accuracy and corresponding overfitting gap as a function of the Hilbert space dimension $K$. At lower dimensions, the full density matrix measurement (FDMM) exhibits a performance advantage over pure-state basis measurement (BM); however, both methods converge to equivalent test accuracies for $K \ge 64$. The overfitting risk, defined as the difference between training and test accuracy, scales with $K$. Results are obtained using $n=4$ discrete processes with evolution times $\tau = [0.2, 0.8, 1.2, 1.4]$.
\textbf{(b)} Sensitivity of single-process test accuracy to the evolution timescale $\tau$, identifying an optimal operational range of $0.1 \le \tau \le 1$. 
\textbf{(c)} Impact of the number of processes $n$ on test accuracy for $K=32$, demonstrating systematic performance enhancement followed by saturation as the feature space expands. \textbf{(d)} Effect of state purity on the full density matrix measurement scheme. As the feature dimension remains invariant with respect to purity, no clear monotonic trend is observed in the classification performance. The times are expressed in units of $10\,$ns.}
    \label{fig:mn1all}
\end{figure*}

\subsection{Representational Diversity}
\label{sec:featureDiversity}

To characterize the underlying structure of the HEF, we first investigate its expressivity by analyzing the dimensionality of the generated feature manifold (details are provided in the Methods~\ref{sec:appendix2}). In Fig.~\ref{fig:mn1}(b), we present the cumulative explained variance obtained via Principal Component Analysis (PCA) of the HEF features. This metric serves as a proxy for the effective rank of the feature space. We identify the minimum number of principal components ($N_{PC}$) required to satisfy standard variance thresholds.

Specifically, at a short evolution time of $\tau=0.2$, we find that $82$, $134$, and $290\pm20$ principal components are necessary to capture $90\%$, $95\%$, and $99\%$ of the total variance, respectively. As the evolution time increases to $\tau=50$, these requirements shift to $94$, $163$, and $356\pm20$ components. This upward trend indicates a spectral broadening of the feature variance; as $\tau$ increases, the information becomes more distributed across the higher-order principal directions. These PCA results suggest that the HEF dynamics project the input data into a nontrivial high-dimensional representation, avoiding both dimensionality collapse and redundant linear expansions.

We further examine the local nonlinearity of the HEF feature map. As illustrated in Fig.~\ref{fig:mn1}(c), the local nonlinearity strength scales monotonically with $\tau$, indicating a progressively higher curvature of the feature manifold at longer evolution times. A high local nonlinearity indicates that the HEF inherently possesses robust, task-agnostic expressivity through strong state-space folding. Taken together with the PCA analysis, these results confirm that larger $\tau$ values generate representations that are simultaneously high-dimensional and geometrically complex.

However, increased expressivity and nonlinearity do not inherently guarantee superior classification performance. To evaluate the utility of these features for supervised learning, we analyze the Fisher ratio in Fig.~\ref{fig:mn1}(d). Since the local nonlinearity is large, as a direct geometric consequence of this initial folding, the untrained Fisher ratio remains relatively small until the linear readout is formally optimized (see Methods~\ref{sec:appendix2}). Interestingly, the untrained Fisher ratio decreases as $\tau$ increases, suggesting that the expanded feature variance does not translate into improved class separability. Instead, the additional degrees of freedom and nonlinearities primarily amplify intra-class fluctuations while providing diminishing returns for inter-class distance. Consequently, while larger $\tau$ processes yield a more expressive reservoir, they may introduce noise-like variations that do not contribute to the discriminative capacity of the model.

These findings clearly distinguish the HEF from purely linear mappings, which are restricted to rotations or scaling of the input coordinates. In contrast to end-to-end trained architectures like Convolutional Neural Networks (CNNs), which actively minimize intra-class spread to achieve neural collapse~\cite{papyan2020neuralcollapse,cohen2020separability}. The HEF utilizes a fixed, high-dimensional dynamical map. The optimal operational regime for the HEF therefore requires a balance between the reservoir's raw expressivity and its untrained class-discriminative structure. We detail the resulting supervised learning performance in the following section.

\subsection{Numerical Results of HEF}
The numerical results presented in Fig.~\ref{fig:mn1all} provide a comprehensive performance analysis of the HEF for the classification of the MNIST data set. Most importantly, our results comprehensively prove that practical utility of a small scale quantum computer can be achieved with an Hilbert space dimension $K\sim 32$ equivalent to a $5$ qubit hardware. The following detailed results provide the numerically obtained test accuracies of the HEF in different configurations and measurement schemes.

Fig.~\ref{fig:mn1all}(a) shows the test accuracy of MNIST data set as a function of Hilbert space dimension $K$. A primary observation from Fig.~\ref{fig:mn1all}(a) is the distinct performance crossover between the two measurement schemes. In low-dimensional regimes, the full density matrix representation provides a marginal advantage in test accuracy because the off-diagonal coherences contribute unique phase information that the basis measurements cannot initially capture. However, as the system scales toward $K \ge 64$, the classification accuracies of both methods converge. This convergence indicates that in sufficiently large Hilbert spaces, the basis measurement alone provides a high-dimensional mapping of the input data that is rich enough to saturate the learning. Consequently, the added complexity of measuring off-diagonal elements becomes redundant.

The study also highlights the critical role of temporal dynamics in feature generation, as seen in Fig.~\ref{fig:mn1all}(b). The identification of an optimal evolution timescale in the range of $0.1 \le \tau \le 1$ suggests a fundamental balance between state separation and information loss. At very short timescales, the system remains too close to its initial state to develop the nonlinearities required for complex feature mapping, whereas excessively long timescales may lead to a loss of input-dependent specificity due to thermalization or decoherence within the reservoir. This temporal sensitivity is further utilized in Fig.~\ref{fig:mn1all}(c), which demonstrates that the systematic concatenation of multiple processes leads to a robust improvement in accuracy. By sampling the reservoir at different points in its evolution, we effectively build a multi-scale representation of the Hamiltonian trajectory, which provides the diversity of features necessary to circumvent the requirement of the physical size of the quantum reservoir (number of qubits).

The sensitivity of the HEF to state purity is shown in Fig.~\ref{fig:mn1all}(d). The results indicate that while the system is weakly responsive to the degree of mixedness, the lack of a clear monotonic trend suggests that the reservoir's performance is not strictly dependent on maintaining absolute state purity. This is particularly promising for experimental realizations where some level of environmental noise is inevitable. Because the feature dimension for full density matrix measurements remains constant regardless of purity, the framework shows a level of robustness that is advantageous for practical hardware implementations. This also implies that the ``informationally complete'' nature of the measurement is more critical than the absolute coherence of the underlying state for the purposes of feature extraction and reservoir computing.

When comparing the efficiency and viability of the two feature reading techniques, the basis-measurement approach emerges as an overall more practical path forward for quantum reservoir processing. At the cost of a moderately larger hardware (Hilbert space dimension), it greatly reduces the measurement complexity, and requiring no complex tomographic post-processing. Since the performance of this hardware-efficient scheme matches the more intensive dense measurement at $K \ge 64$, it offers a way to scale quantum reservoir computing without incurring the quadratic sample complexity or the high gate error rates associated with full state tomography. Therefore, the basis measurement-based feature extraction method is not only more efficient in terms of measurement resources but also more promising for quantum hardware with limited coherence times where minimizing the readout overhead is essential for maintaining high fidelity.

%%%%%%%%%%%
\begin{figure}
    \centering
    \includegraphics[width=0.95\linewidth]{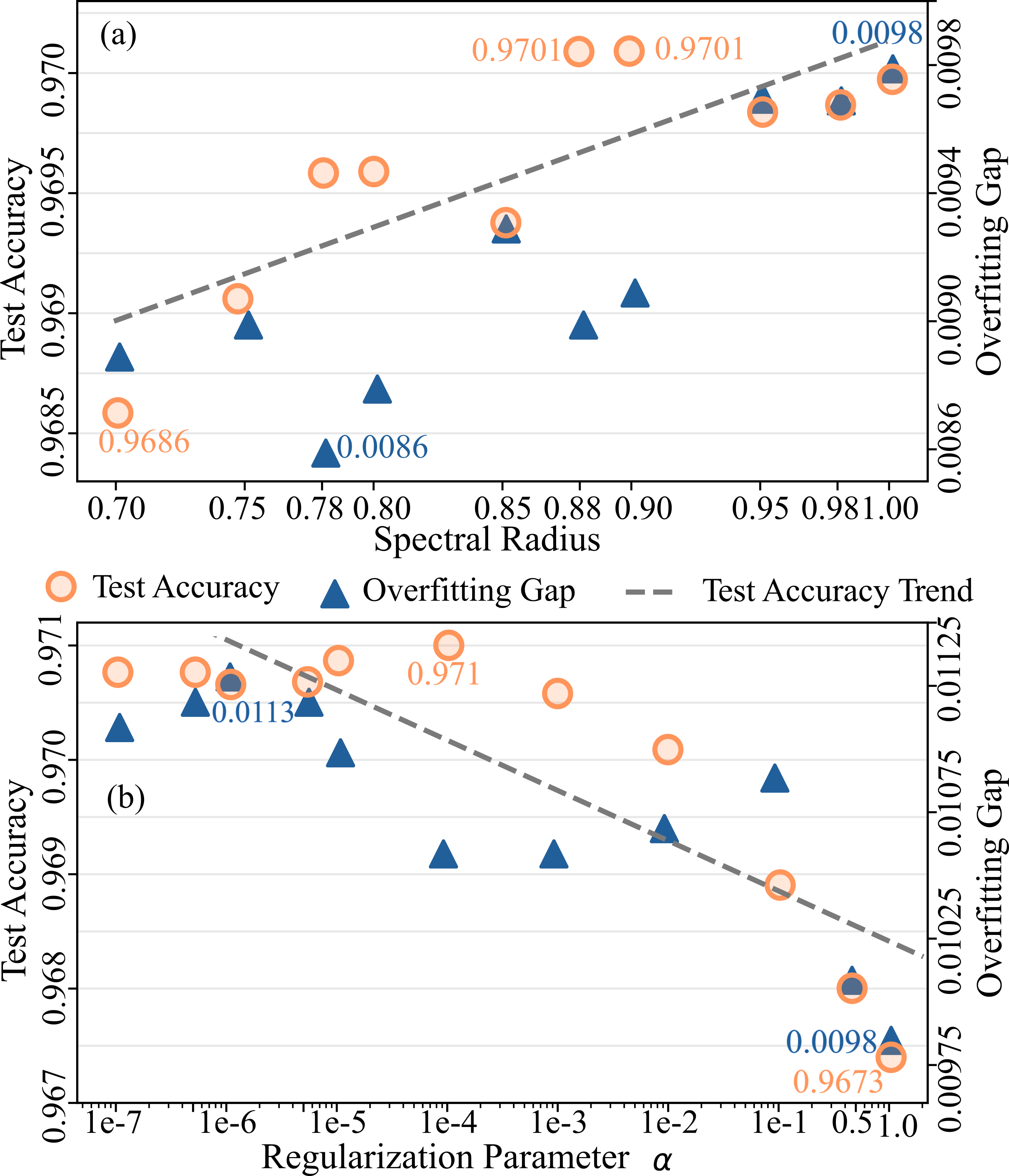}
   \caption{\textbf{Hyperparameter assessment for the HEF.} \textbf{(a)} Impact of the spectral radius on classification performance. The test accuracy exhibits an increasing trend as the spectral radius approaches unity, whereas the overfitting gap remains relatively stable within a narrow interval. \textbf{(b)} Influence of the regularization parameter $\alpha$ on model generalization. The test accuracy shows a marginal, gradual decline as $\alpha$ is increased across several orders of magnitude (from $10^{-7}$ to $1$), while the overfitting gap demonstrates only minor fluctuations across the sampled range.}
    \label{fig:mn1train}
\end{figure}

\subsection{Hyperparameter Stability}
 
Hyperparameter selection is an important part of training in learning process, since the final performance often depends not only on the model structure itself but also on the choice of global control parameters~\cite{bergstra2012random,yang2020hyperparameter}. In many machine learning frameworks, inappropriate hyperparameter settings may lead to overfitting, underfitting, or unstable performance~\cite{goodfellow2016regularization}. For the proposed framework, two particularly important global hyperparameters are the regularization parameter $\alpha$ in the classical readout and the spectral radius $r_\text{tr}$ that controls the effective scale of the reservoir Hamiltonian.

The regularization parameter $\alpha$ determines the penalty strength within the classical linear classifier, thereby mediating the trade-off between minimizing training error and ensuring model generalization~\cite{scikit_ridge}. While a sufficiently small $\alpha$ allows for high-fidelity fitting of the reservoir features, an excessively large value may suppress critical discriminative information. Conversely, the spectral radius dictates the global dynamical scale of the reservoir Hamiltonian, directly influencing the stability and richness of the emergent features~\cite{jaeger2007scholarpedia,lukosevicius2012practical,venayagamoorthy2009spectral}. Given that the quality of the learned representation is intrinsically tied to the reservoir's underlying dynamics, the spectral radius serves as a primary metric for assessing the sensitivity of the feature generation process to global scaling.

To evaluate the influence of these parameters, we conducted a systematic sensitivity analysis by varying each hyperparameter independently. The spectral radius was sampled across the interval $[0.7, 1.0]$, while the regularization parameter $\alpha$ was swept across several orders of magnitude ($10^{-7}$ to $10^0$) to encompass both weak and strong regularization regimes. The results of this analysis are summarized in Fig.~\ref{fig:mn1train}.

As illustrated in Figs.~\ref{fig:mn1train}(a) and (b), both the test accuracy and the overfitting gap exhibit remarkably low sensitivity to these variations. Specifically, the observed fluctuations remain within an order of $10^{-4}$, demonstrating that the HEF architecture is highly stable with respect to its global training parameters. This inherent stability suggests that the framework does not require extensive hyperparameter fine-tuning to achieve optimal performance, further underscoring its robustness and practical utility for quantum reservoir processors.

\section{Analog Superconducting Array Processor (ASAP)}\label{sec:theory2}
 
\begin{figure*}
    \centering
    \includegraphics[width=0.95\linewidth]{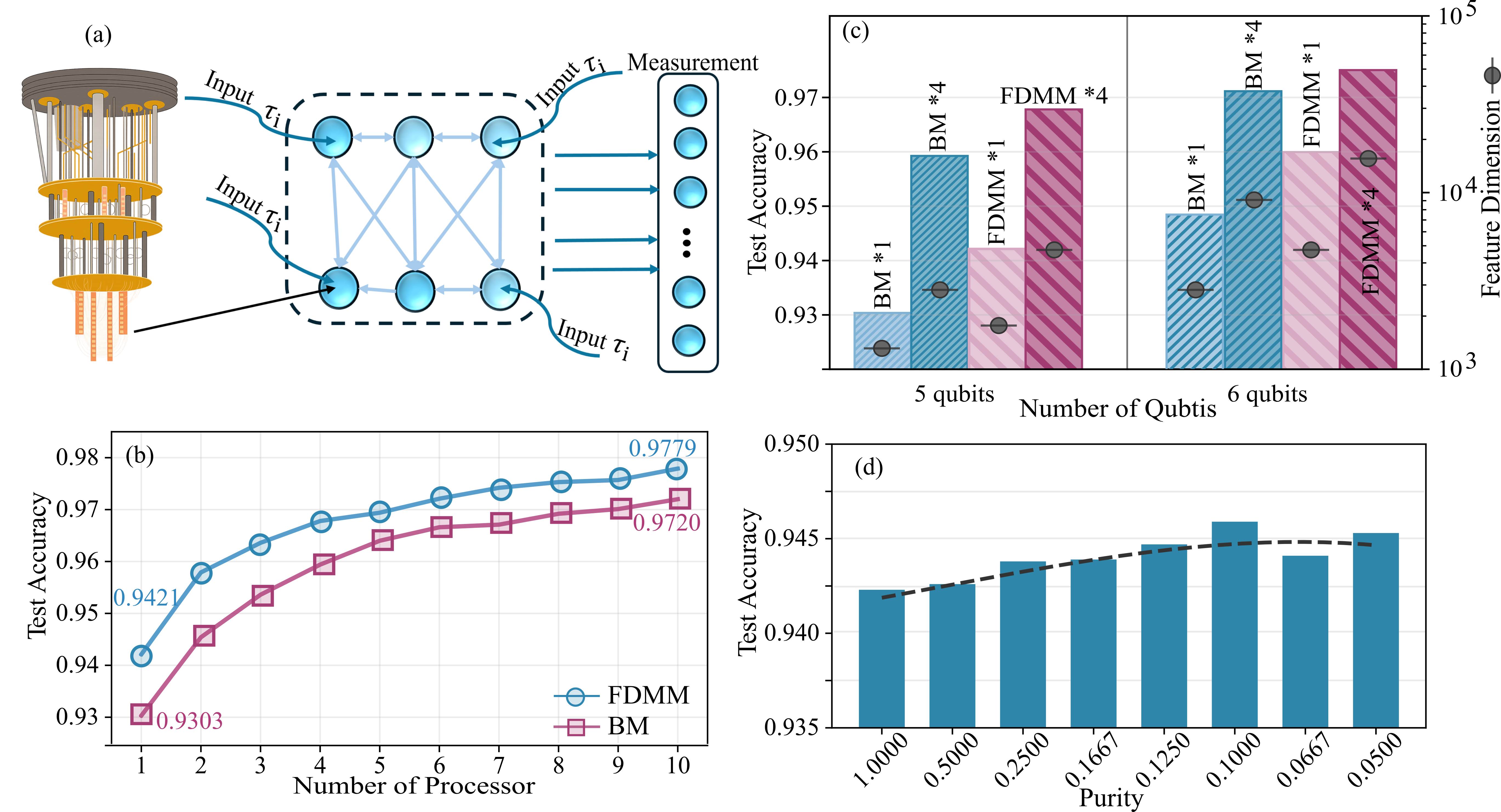}
    \caption{
    \textbf{Analog Superconducting Array Processor (ASAP) architecture and performance benchmarks.} \textbf{(a)} Schematic representation of the ASAP system. Input data are mapped into a superconducting qubit array via independent control channels, followed by analog quantum evolution. The resulting state is sampled through either basis or informationally complete measurements to generate high-dimensional features for subsequent classification. \textbf{(b)} Scaling performance with respect to the number of  ASAP processes. The test accuracy is plotted for increasing numbers of parallel processors utilizing distinct evolution timescales for both full density matrix measurement (FDMM) and basis measurement (BM) schemes. \textbf{(c)} Influence of qubit count on classification accuracy and feature space dimensionality. Test accuracy is shown as a function of the number of qubits for both FDMM and BM schemes. The secondary axis (gray markers) indicates the corresponding feature dimensions on a logarithmic scale. As indicated in the figure, the accuracies are obtained for $1$ and $4$ processes. \textbf{(d)} Sensitivity of the full density matrix implementation to state purity. The results demonstrate that classification performance remains robust to variations in the mixed-state purity. We observe a marginal improvement in classification performance as purity decreases.}
    %\new{BM: Basis Measurement, FDMM: Full Density Matrix Measurement. *1: one process, *4: four process.}
    \label{fig:mn2all}
\end{figure*}

For the physical realization of the HEF, we propose a platform based on an array of interconnected superconducting qubits operating as an analog processor. Unlike the conventional digital circuit model, this approach bypasses the requirement for discrete quantum gate operations. Instead, it uses the natural many-body interactions within a superconducting network to dynamically generate high-dimensional features. We call this realization as the Analog Superconducting Array Processor (ASAP), as illustrated in Fig.~\ref{fig:mn2all}(a).

From the perspective of utility, the ASAP architecture demonstrates significant near-term feasibility. Because the QRP framework is functionally independent of specific Hamiltonian structures and the HEF encoding scheme is highly flexibility, the many-body Hamiltonian of typical superconducting qubit arrays can be directly utilized for ASAP applications. Superconducting systems offer a well-established platform for realizing coherent quantum dynamics characterized by a high degree of controllability~\cite{devoret2013superconducting, krantz2019quantum}.

We consider an effective superconducting Hamiltonian, $H_{\mathrm{SC}}$, which captures the coherent processes in circuit quantum electrodynamics (cQED)~\cite{blais2021circuit}:
\begin{equation}
\label{eq:sc_hamiltonian}
\begin{aligned}
H_{\mathrm{SC}} =\;&
\sum_{j=1}^{N_q} \frac{\omega_j}{2}\,\sigma_j^z
+ \sum_{i<j} J_{ij}
\left(
\sigma_i^+ \sigma_j^- + \sigma_i^- \sigma_j^+
\right) \\
&+ \sum_{j=1}^{N_q}
\left(
P_j^x \sigma_j^x + P_j^y \sigma_j^y
\right).
\end{aligned}
\end{equation}
Here, $N_q$ denotes the number of qubits, and $\omega_j$ represents the energy splitting of the $j$-th qubit. The term $J_{ij}$ characterizes the exchange interaction strength between qubits $i$ and $j$, while $P_j^x$ and $P_j^y$ denote the amplitudes of coherent microwave drives applied along the $x$ and $y$ quadratures, respectively. The operators $\sigma_j^x$, $\sigma_j^y$, and $\sigma_j^z$ are the standard Pauli matrices. To bridge this theoretical Hamiltonian with the practical operation of the ASAP, we detail its numerical implementation in the following section.

Two-qubit interactions in Eq.~\eqref{eq:sc_hamiltonian} are modeled via excitation exchange terms, which naturally arise from capacitive coupling or interactions mediated by a shared microwave resonator in cQED. Here, superconducting qubits couple to each other through the exchange of quantized microwave photons. We construct the coupling strength matrix $J_{ij}$ as a weak, dense interaction matrix, yielding a nonlocal exchange term that promotes extensive state mixing across the reservoir. The specific form of the coupling is given by:
\begin{equation}
J_{ij} = \frac{g_i g_j}{\Delta_{ij}} + \delta J_{ij},
\end{equation}
where $g_i$ denotes the effective coupling strength between qubit $i$ and the shared resonator or bus mode, and $\Delta_{ij}$ represents the effective detuning governing the second-order virtual exchange process. The term $\delta J_{ij}$ accounts for weak residual direct couplings and hardware imperfections inherently present in superconducting platforms~\cite{blais2004cavity}. Detailed numerical configurations for these parameters are provided in Methods~\ref{appendix1}.

%%%%%%%%%%%%%%%%%%%%%%%%%%%%%%
\subsection{ASAP Data Encoding}\label{ASAPDataEncoding}
%%%%%%%%%%%%%%%%%%%%%%%%%%%%%%
In the ASAP, input data is encoded into the Hamiltonian $H_{\mathrm{SC}}$ through the parameters $\omega_j$, $P_j^x$ and $P_j^y$. The single-qubit terms $\sum_j (\omega_j/2)\,\sigma_j^z$ represent the effective transition frequencies of the qubits. To implement the HEF framework in conjunction with the QRP, these frequency parameters $\omega_j$ are constructed as a hybrid of data-dependent and stochastic components. Specifically, the classical input is first linearly encoded and partitioned into sequential chunks; for each $n$-th chunk, the corresponding frequency vector is defined as
\begin{equation}\label{eq:encodew}
\boldsymbol{\omega}^{(n)} = \delta \boldsymbol{\omega} + W_{\omega}\,\mathbf{x}_{\mathrm{chunk}}^{(n)},
\end{equation}
where $\mathbf{x}_{\mathrm{chunk}}^{(n)}$ denotes the encoded classical input components for the $n$-th interval, $W_{\omega}$ is a linear mapping matrix, and $\delta \boldsymbol{\omega}$ represents static frequency offsets drawn from a normal distribution, accounting for intrinsic device disorder or calibration uncertainties in the data-input interface. 

Coherent external control is integrated through the transverse driving terms $\sum_j\left( P_j^x \sigma_j^x + P_j^y \sigma_j^y \right)$, which correspond to the in-phase and quadrature microwave drives applied to the qubit array. In our numerical simulations, the drive amplitudes $(P_j^x, P_j^y)$ are derived from a linear transformation of the input chunks combined with static bias terms. For the $n$-th encoded input chunk $\mathbf{x}_{\mathrm{chunk}}^{(n)}$, the transverse drive amplitudes are constructed as
\begin{equation}\label{eq:encodepxpy}
\mathbf{P}_x^{(n)} = W_x \mathbf{x}_{\mathrm{chunk}}^{(n)} + \mathbf{b}_x,
\qquad
\mathbf{P}_y^{(n)} = W_y \mathbf{x}_{\mathrm{chunk}}^{(n)} + \mathbf{b}_y,
\end{equation}
where $W_x$ and $W_y$ represent the linear mapping matrices that translate classical data into microwave drive strengths, while $\mathbf{b}_x$ and $\mathbf{b}_y$ are small static bias vectors that ensure the system operates within a stable dynamical regime. This formulation allows the ASAP to map classical information directly into the Hamiltonian parameters, thereby driving the complex quantum evolution required for feature generation.

%%%%%%%%%%%%%%%%%%%%%%%%%%%
\subsection{Analog Superconducting Dynamics}\label{sec:scdynamics}
 
Since we partition the input data into sequential chunks, each complete evolution cycle must systematically integrate the information from all chunks. We implement this by decomposing the global time evolution into a series of sequential sub-evolutions. For a given timescale $\tau$, the encoded input sequence $\{\mathbf{x}_{\mathrm{chunk}}^{(n)}\}_{n=1}^{N_c}$ generates a corresponding sequence of chunk-dependent Hamiltonians $\{H_n\}_{n=1}^{N_c}$.

For a fixed evolution timescale $\tau$, each chunk induces an analog unitary evolution defined as
\begin{equation}
U_n(\tau)=e^{-iH_n\tau}, \qquad n=1,2,\dots,N_c,
\end{equation}
The total evolution operator at time $\tau$ is then given by the time-ordered product
\begin{equation}
\label{eq:total_unitary_chunk}
U(\tau)=U_{N_c}(\tau)\cdots U_2(\tau)U_1(\tau),
\end{equation}
which preserves the sequential injection of the encoded classical information throughout the chunked control structure. Starting from a fixed initial state $\rho_0$, the resulting output density matrix is
\begin{equation}
\rho(\tau)=U(\tau)\rho_0U^\dagger(\tau).
\end{equation}
To enrich the feature representation, we execute multiple processes across a set of distinct evolution timescales $\{\tau_1,\tau_2,\dots,\tau_m\}$. For each $\tau_i$, an output density matrix $\rho(\tau_i)$ is obtained independently. The final feature vector is constructed by concatenating the measurements---utilizing either the basis measurement or full density matrix measurement schemes described in Sec.~\ref{sec:heffeatures}---across all selected timescales. This multi-scale approach ensures that the reservoir captures both fast and slow dynamical correlations within the input data.

\subsection{Numerical Results of ASAP}

We evaluate the effectiveness of the ASAP platform governed by the effective Hamiltonian in Eq.~\eqref{eq:sc_hamiltonian}. The performance analysis is conducted using the full MNIST dataset with a five-qubit ASAP configuration. Fig.~\ref{fig:mn2all} illustrates that the ASAP serves as an effective platform for implementing the QRP through the HEF scheme, achieving classification accuracies as high as $98\%$ with only a five-qubit processor.

Fig.~\ref{fig:mn2all}(b) provides a systematic investigation into the influence of the number of processes on a five-qubit processor. An increase in the number of ASAP processes yields a progressive improvement in test accuracy, which is accompanied by a proportional expansion of the feature dimensionality. This gain eventually enters a saturation regime as further timescales are incorporated. Such a trend indicates that incorporating multiple processes with varied temporal evolutions enriches the dynamical representation while maintaining a controlled increase in complexity. Notably, the basis measurement scheme achieves a test accuracy nearly identical to that of the full density matrix measurement scheme, further validating the utility of basis measurements for the ASAP architecture.

In Fig.~\ref{fig:mn2all}(c), the classification performance is compared between five-qubit and six-qubit ASAP configurations. The test accuracy increases marginally with the qubit count, primarily as a result of the larger number of extracted features. While both measurement schemes exhibit performance gains as the system size expands, the density matrix representation consistently maintains superior accuracy across all configurations. It is significant to note that while the feature dimensionality grows rapidly with $N_q$, the improvement in test accuracy remains incremental. This suggests that a five-qubit ASAP already generates sufficient feature diversity for the task, rendering additional features largely redundant.

The impact of state purity on classification is depicted in Fig.~\ref{fig:mn2all}(d). The test accuracy remains stable across a wide range of purity values, from 1 down to 0.05. The ASAP demonstrates no systematic performance degradation, implying that moderate levels of mixedness do not disrupt the underlying reservoir dynamics. Furthermore, the observation that moderate mixedness may marginally improve performance suggests an implicit regularizing effect. 

Collectively, these findings demonstrate that the behaviors identified in the abstract HEF model are preserved when the dynamics are constrained by a physically motivated superconducting Hamiltonian. The ASAP exhibits systematic scaling regarding both system size and temporal depth and remains robust under mixed-state conditions.

%%%%%%%%%%%%%%%%%%%%%%
\section{Quantum Circuit Implementation (QCI)}\label{sec:theory3}

We present a framework for realizing the HEF scheme via shallow digital quantum circuits, designated as the Quantum Circuit Implementation (QCI). This digital formulation facilitates the emulation of reservoir dynamics through discrete sequences of quantum gates, providing a versatile and scalable architecture for feature generation.

The circuit architecture is partitioned into two functional modules: (i) the HEF block and (ii) the interaction block. The HEF block is responsible for data encoding and performing nonlinear transformations through localized single-qubit rotations. Conversely, the interaction block introduces coherent coupling between the qubits, thereby establishing the necessary interconnections to facilitate many-body state mixing across the reservoir.

\begin{figure*}
    \centering
    \includegraphics[width=\linewidth]{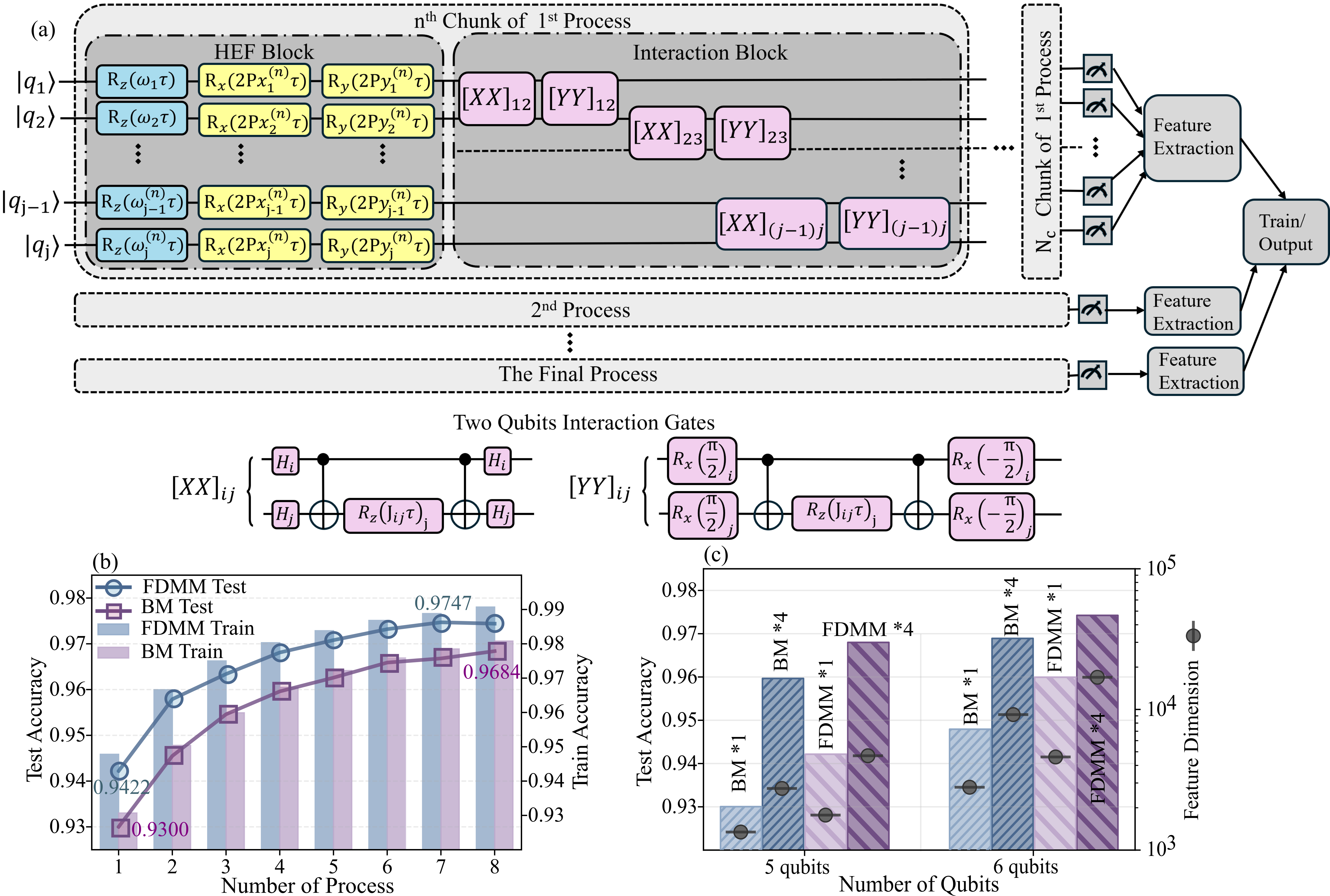}
    \caption{\textbf{Quantum Circuit Implementation (QCI) architecture and performance bechmarks.}
    \textbf{(a)} The input is encoded by using the similar strategy of HEF, then inject the encoded data into the $1^{st}$ process, measurement outcomes are collected for feature extraction, providing a gate based implementation of HEF. Then, restart the experiment to the next process, until the final process, finally concentrate all features for training. Inside each process, QCI includes encoding plus nonlinear transformation block and intersection block. The two qubits interaction gates are realized under many body Hamiltonian through $XX$ and $YY$ interaction blocks, shown at the lower side.
    \textbf{(b)} Test accuracy for increased number of QCI process by using different time scale for both full density matrix measurement (FDMM) and basis measurement (BM) schemes.
    \textbf{(c)} Influence of qubit count on classification accuracy and feature space dimensionality. Test accuracy is shown as a function of the number of qubits for both FDMM and BM schemes. The secondary axis (gray markers) indicates the corresponding feature dimensions on a logarithmic scale.  As indicated in the figure, the accuracies are obtained for $1$ and $4$ processes.}
    %\new{BM: Basis Measurement, FDMM: Full Density Matrix Measurement. *1: one process, *4: four process.}
    \label{fig:mn3_1}
\end{figure*}

\subsection{HEF Block}
In the HEF block of the quantum circuit, classical information is encoded via single-qubit rotations. The input data is partitioned into sequential chunks and mapped to Hamiltonian parameters according to the linear encoding rules defined in Eqs.~\eqref{eq:encodew} and \eqref{eq:encodepxpy}. For the $n$-th chunk, the unitary evolution of the $j$-th qubit is represented by a sequence of Euler rotations:
\begin{equation}
U_\text{HEF}^{(n),j}(\tau) = R_z(\omega_j^{(n)} \tau) R_x(2P_{x_j}^{(n)} \tau) R_y(2P_{y_j}^{(n)} \tau).
\end{equation}
The collective HEF unitary for the $n$-th chunk is the tensor product of these single-qubit operations across all the qubits:
\begin{equation}
U_\text{HEF}^{(n)}(\tau) = \bigotimes_{j=1}^{N_q} U_\text{HEF}^{(n),j}(\tau).
\end{equation}
At this stage establishing interconnections between the qubits is required. In the absence of qubit interconnections, the application of sequential single-qubit rotations for successive data chunks is mathematically insufficient for high-dimensional feature representation. Because any sequence of local rotations---regardless of whether they act along the $x$, $y$, or $z$ axes---is equivalent to a single effective rotation within the $SU(2)$ manifold, the system cannot independently preserve the information contained within discrete, sequential chunks. Consequently, the introduction of interconnections between qubits after each data injection is essential to facilitate entanglement and many-body mixing. This process ensures that the temporal identity of each chunk is mapped into a larger, non-local Hilbert space, preventing the information from being compressed into a single effective local operation.

\subsection{Interaction Block}
The interaction block consists of two-qubit operations that establish coherent interconnections with fixed pairwise coupling strengths $J_{jk}$. This block is applied following the injection of each input chunk. The unitary for a given qubit pair $(j, k)$ is defined by:
\begin{equation}
U^{j,k}_\text{INT}(\tau) = e^{-\frac{i\tau}{2} J_{jk} X_j X_{k} } e^{-\frac{i\tau}{2} J_{jk} Y_j Y_{k}}.
\end{equation}
The $XX$ sector of the interaction is implemented via basis transformations and a central $R_z$ rotation:
\begin{align}
e^{-\frac{i\tau}{2} J_{ij} X_i X_j} = (H_i \otimes H_j) &\text{CNOT}_{ij}[R_z(J_{ij}\tau)]_j  \notag\\ &\text{CNOT}_{ij} (H_i \otimes H_j),
\end{align}
where $H_i$ and $\text{CNOT}_{ij}$ denote the Hadamard and Controlled-NOT gates, respectively. Similarly, the $YY$ sector is realized by rotating the qubits into the $Y$-basis:
\begin{align}
e^{-iJ_{ij} Y_i Y_j \tau/2} = [R_x(\tfrac{\pi}{2})]_{ij} &\text{CNOT}_{ij} [R_z(J_{ij}\tau)]_j\notag\\ 
& \text{CNOT}_{ij} [R_x(-\tfrac{\pi}{2})]_{ij}.
\end{align}
This interaction block is applied exclusively to the nonzero coupling edges of the interaction graph, thereby respecting the physical connectivity of the hardware. The total interaction unitary is given by the ordered product:
\begin{equation}
U_\text{INT} (\tau)= \prod_{j=1, k>j}^{N_q} U^{j,k}_\text{INT}(\tau).
\end{equation}
Here, $k$ runs only over those indices with non-zero coupling constants $J_{jk}$.

\subsection{Full Unitary and Feature Extraction}
The evolution associated with a single input chunk is defined by the combined unitary $U^{(n)}(\tau)$:
\begin{equation}
U^{(n)}(\tau) = U_\text{INT}(\tau) U^{(n)}_\text{HEF}(\tau).
\end{equation}
To process the complete sequence of $N_c$ chunks, the system undergoes a total unitary evolution $U_\text{QIC}(\tau)$, represented as a time-ordered product:
\begin{equation}
U_\text{QIC}(\tau)= \prod_{n=1}^{N_c} U^{(n)}(\tau) = U^{(N_c)}(\tau) \dots U^{(2)}(\tau) U^{(1)}(\tau).
\end{equation}
Given an initial state $\rho_0$, the output density matrix is obtained as $\rho(\tau) = U_\text{QIC}(\tau) \rho_0 U_\text{QIC}^\dagger(\tau)$. Features are subsequently extracted using either the full density matrix or basis measurement schemes as detailed in Sec.~\ref{sec:heffeatures}.

\subsection{Numerical Results of QCI}
We numerically benchmark the Quantum Circuit Implementation (QCI) using the full MNIST dataset, with results presented in Figs.~\ref{fig:mn3_1}(b) and \ref{fig:mn3_1}(c). A defining characteristic of the QCI is its shallow circuit depth, which scales linearly with the input data dimensionality $N_c$. This limited depth is particularly advantageous for experimental realization on current quantum hardware, where gate operations must be completed within the constraints of finite qubit coherence times.

Fig.~\ref{fig:mn3_1}(b) illustrates the impact of the number of QCI processes on a five-qubit processor. As the process count increases from $1$ to $8$, both measurement strategies exhibit a systematic improvement in test accuracy. The consistently superior performance of the full density matrix measurement indicates that this higher-dimensional representation captures a broader range of dynamical correlations compared to the basis measurement scheme. While the training accuracy approaches $\sim 98\%$, the saturation of the test accuracy at higher process counts suggests that additional processes primarily enhance the model's expressive power, while the marginal gains in generalization begin to diminish.

The classification performance for five-qubit and six-qubit configurations is summarized in Fig.~\ref{fig:mn3_1}(c). Increasing the number of qubits results in a marginal performance gain, which remains stable across different process counts. This improvement is primarily attributed to the increased diversity of the generated feature set. However, the incremental nature of the accuracy increase relative to the growth in system size suggests that a five-qubit QCI already provides sufficient dynamical complexity to represent the MNIST features effectively, rendering further expansion largely redundant for this specific task.

\section{Comparison: ASAP vs. QCI}\label{sec:comparison}
We have evaluated the proposed HEF using two distinct realization paths: the Analog Superconducting Array Processor (ASAP) and the Quantum Circuit Implementation (QCI), as detailed in Secs.~\ref{sec:theory2} and \ref{sec:theory3}. Both architectures demonstrate consistently high classification performance within a low-qubit regime, achieving accuracies exceeding 97\% with a five-qubit configuration and an incremental gain of approximately 1\% upon transitioning to six qubits. 

The framework exhibits exceptional stability; performance fluctuations remain on the order of $10^{-3}$ when varying critical parameters such as state purity (under full density matrix measurements) or the number of temporal multiplexing processes. This robustness indicates that high-fidelity feature extraction is achievable across diverse physical implementation paradigms. Significantly, these results suggest that competitive performance is not restricted to digital gate-based systems but is equally attainable through analog superconducting methods. Given that digital implementations often incur significant computational overhead due to gate decomposition and error accumulation, the analog approach emerges as a highly resource-efficient alternative for practical hardware realization.

Furthermore, the HEF architecture effectively addresses a primary challenge in variational quantum neural networks (QNNs): the \textit{barren plateau} problem. By utilizing a fixed quantum reservoir and restricting training to a classical linear layer, the framework avoids the vanishing gradient issues that typically hinder convergence and performance in traditional quantum machine learning models, as discussed in Sec.~\ref{sec:theory1}.

\section{Dissipation}\label{sec:opensystem}
%%%%%%%%%%%%%%%%%%%%%%%%%%%%%%%
In the presence of environmental coupling, the generalized quantum evolution of a system of qubits is governed by the Lindblad master equation~\cite{lindblad1976}, which provides a Markovian description of the density matrix $\rho$:
\begin{equation}
\label{eq:lindblad_general}
\frac{d\rho}{dt}
=
-i[H_\text{SC},\rho]
+
\sum_k
\gamma_k
(
L_k \rho L_k^\dagger
-
\frac{1}{2}
\{
L_k^\dagger L_k,\rho
\})
\end{equation}
where $\{\cdot,\cdot\}$ denotes the anti-commutator. The operators $L_k$ are the Lindblad (or collapse) operators that define the specific channels of interaction with the environment, while the coefficients $\gamma_k \ge 0$ represent the associated relaxation or decoherence rates.

The first term on the right-hand side of Eq.~\eqref{eq:lindblad_general}, represents the unitary part of the evolution. In the limit of an isolated system $(\gamma_k=0)$, this term induces the von Neumann evolution, preserving the purity of the quantum state. The second term, referred to as the Lindbladian dissipator, describes the non-unitary and irreversible influence of the environment. 

To account for decoherence in the ASAP configuration, we incorporate two primary dissipative channels within the Lindblad formalism. The first is pure dephasing, which models the loss of phase coherence without inducing transitions between energy levels. This process is described by the collapse operators $L_j^{(\phi)} = \sigma_j^z$ with a corresponding dephasing rate $\gamma_\phi$. 

The second channel is amplitude damping, representing the energy relaxation of the qubits from the excited state to the ground state. This is governed by the lowering operators $L_j^{(1)} = \sigma_j^-$ with a relaxation rate $\gamma_1$. By incorporating these channels into the general Markovian master equation, the time evolution of the system's density matrix $\rho$ is given by:
\begin{equation}
\label{eq:lindblad_specific}
\begin{aligned}
\frac{d\rho}{dt} = & -i[H_\text{SC},\rho] + \sum_{j=1}^{N_q} \gamma_\phi \left( \sigma_j^z \rho \sigma_j^z - \rho \right) \\
& + \sum_{j=1}^{N_q} \gamma_1 \left( \sigma_j^- \rho \sigma_j^+ - \frac{1}{2} \left\{ \sigma_j^+\sigma_j^-, \rho \right\} \right)
\end{aligned}
\end{equation}
In the dephasing term, we have utilized the identity $(\sigma_j^z)^2 = \mathbf{1}$ to simplify the dissipator.

The master equation [Eq.~\eqref{eq:lindblad_specific}] can be formally vectorized and recast in Liouville space as $\dot{\vec{\rho}} = \mathcal{L}\vec{\rho}$, where $\vec{\rho}$ denotes the vectorized density matrix and $\mathcal{L}$ represents the Liouvillian superoperator. Within this representation, the open-system evolution over a 

time interval $\tau$ is governed by the propagator $\mathcal{S}(\tau) = e^{\tau \mathcal{L}}$, such that:
\begin{equation}
\vec{\rho}(\tau) = e^{\tau \mathcal{L}} \vec{\rho}_0.
\end{equation}

For a sequence consisting of $N_c$ data chunks, each input determines a specific Hamiltonian $H_n$, which in turn defines a corresponding chunk-dependent Liouvillian $\mathcal{L}_n$. The evolution of the system is thus described by a composition of these dynamical maps. The final output state is obtained from the time-ordered product of the $N_c$ individual evolutions:
\begin{equation}
\vec{\rho}_{\mathrm{out}} = e^{\tau\mathcal{L}_{N_c}} \cdots e^{\tau\mathcal{L}_2} e^{\tau\mathcal{L}_1} \vec{\rho}_0.
\end{equation}

Following this sequential evolution, the ASAP features are extracted from the elements of $\vec{\rho}_{\mathrm{out}}$ for subsequent classical processing. This Liouvillian formulation provides a computationally efficient framework for simulating the reservoir dynamics.

For the QCI framework, we treat the digital circuit as an open quantum system by preserving its gate-decomposed structure while incorporating discrete noise channels. Rather than performing a continuous-time integration of the master equation, we emulate the physical dissipation and dephasing mechanisms by interspersing quantum noise maps after each unitary chunk $U_n(\tau)$. Specifically, for a given chunk-dependent operation, the system density matrix $\rho$ is updated according to:
\begin{equation}
\label{eq:qci_noisy_update}
\rho \;\mapsto\; \mathcal{E}_{\mathrm{loss}} \left( U_n(\tau)\,\rho\,U_n^\dagger(\tau) \right),
\end{equation}
where $\mathcal{E}_{\mathrm{loss}}$ represents a composite error channel acting independently on all qubits. In this formulation, $\mathcal{E}_{\mathrm{loss}}$ is composed of amplitude damping and phase damping channels, providing a discrete-time analog to the longitudinal relaxation and dephasing processes described in Eq.~\eqref{eq:lindblad_specific}. 

Utilizing the Kraus representation, the action of a general quantum channel is expressed as:
\begin{equation}
\label{eq:kraus_general}
\mathcal{E}(\rho)=\sum_{\mu} K_\mu \rho K_\mu^\dagger, \qquad \sum_{\mu} K_\mu^\dagger K_\mu = I.
\end{equation}
For the amplitude damping channel, the single-qubit Kraus operators are:
\begin{equation}
\label{eq:amp_kraus}
K_0^{(1)}= \begin{pmatrix} 1 & 0\\ 0 & \sqrt{1-p_1} \end{pmatrix}, \qquad K_1^{(1)}= \begin{pmatrix} 0 & \sqrt{p_1}\\ 0 & 0 \end{pmatrix},
\end{equation}
while the phase damping channel is characterized by:
\begin{equation}
\label{eq:phase_kraus}
K_0^{(\phi)}= \begin{pmatrix} 1 & 0\\ 0 & \sqrt{1-p_\phi} \end{pmatrix}, \qquad K_1^{(\phi)}= \begin{pmatrix} 0 & 0\\ 0 & \sqrt{p_\phi} \end{pmatrix}.
\end{equation}
Here, $p_1$ and $p_\phi$ denote the cumulative decay probabilities over the chunk duration $\tau$. To maintain physical consistency with the continuous-time rates defined in Eq.~\eqref{eq:lindblad_specific}, we parameterize these probabilities based on an exponential Markovian decay model. The survival probability over the interval $\tau$ is given by $e^{-\gamma \tau}$, leading to the mapping:
\begin{equation}
\label{eq:rate_to_prob}
p = 1 - e^{-\gamma \tau},
\end{equation}
where $\gamma = \gamma_1$ for amplitude damping and $\gamma = \gamma_\phi$ for phase damping~\cite{breuer2002theory}.

\begin{figure}
    \centering
    \includegraphics[width=0.95\linewidth]{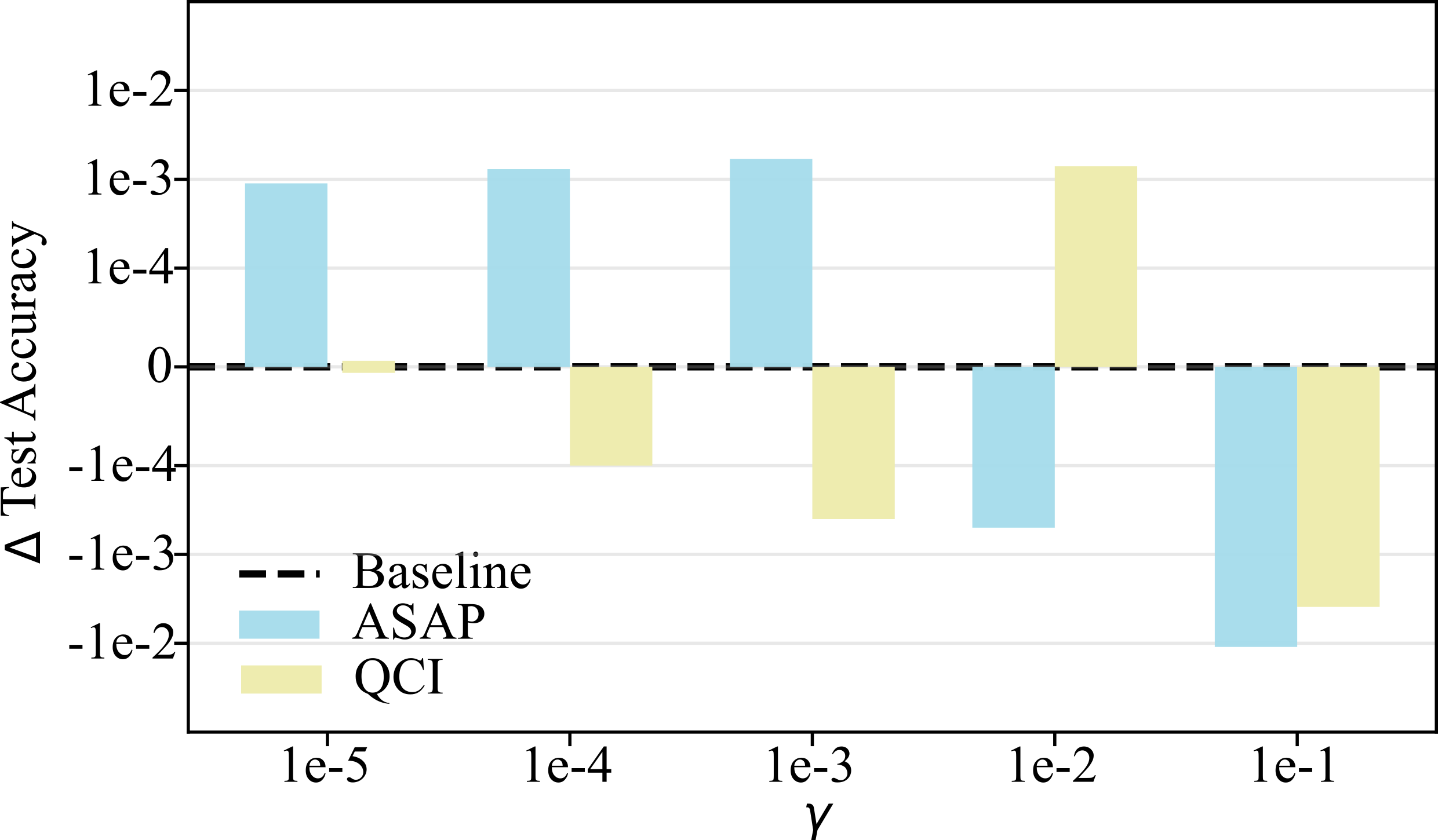}
    \caption{\textbf{Effect of dissipation on the ASAP and QCI at short evolution times.} The accuracy deviation $\Delta$, evaluated as the difference between the test accuracies with and without dissipation, is plotted against the dissipation rate $\gamma$. Over a wide range of dissipation rates spanning $10^{-5}$ to $10^{-1}$, the accuracy deviation remains bounded around $10^{-3}$ (corresponding to a minor $0.1\%$ performance shift). This demonstrates the excellent robustness of the HEF-based QRP.}
 \label{fig:mn23loss}
\end{figure}

\textit{Small $\tau$ regime:--} In the small $\tau$ regime ($\tau \in \{0.2, 0.8\}$), dissipative effects are largely negligible. Figure~\ref{fig:mn23loss} summarizes the performance of 5-qubit ASAP and QCI implementations relative to their dissipation-free baselines [Figs.~\ref{fig:mn2all}(c), \ref{fig:mn3_1}(b)], with relaxation rates $\gamma = \gamma_1 = \gamma_{\phi}$ varied from $10^{-5}$ to $10^{-1}$. Despite the short chunk duration, the aggregate evolution time ($N_c \tau > 10$) ensures the system undergoes significant dynamical interaction. Both methods exhibit nearly identical accuracy trends, with noise-induced deviations remaining below $10^{-3}$, demonstrating the framework's robustness under realistic environmental coupling. While minor performance dips reflect the loss of coherence, occasional marginal improvements suggest that dissipation may help mitigate quantum scrambling~\cite{Xiong2025}---a phenomenon driven by accumulated coherent oscillations that typically intensifies as the system size or evolution time scales up.

\begin{figure}
     \centering
     \includegraphics[width=1\linewidth]{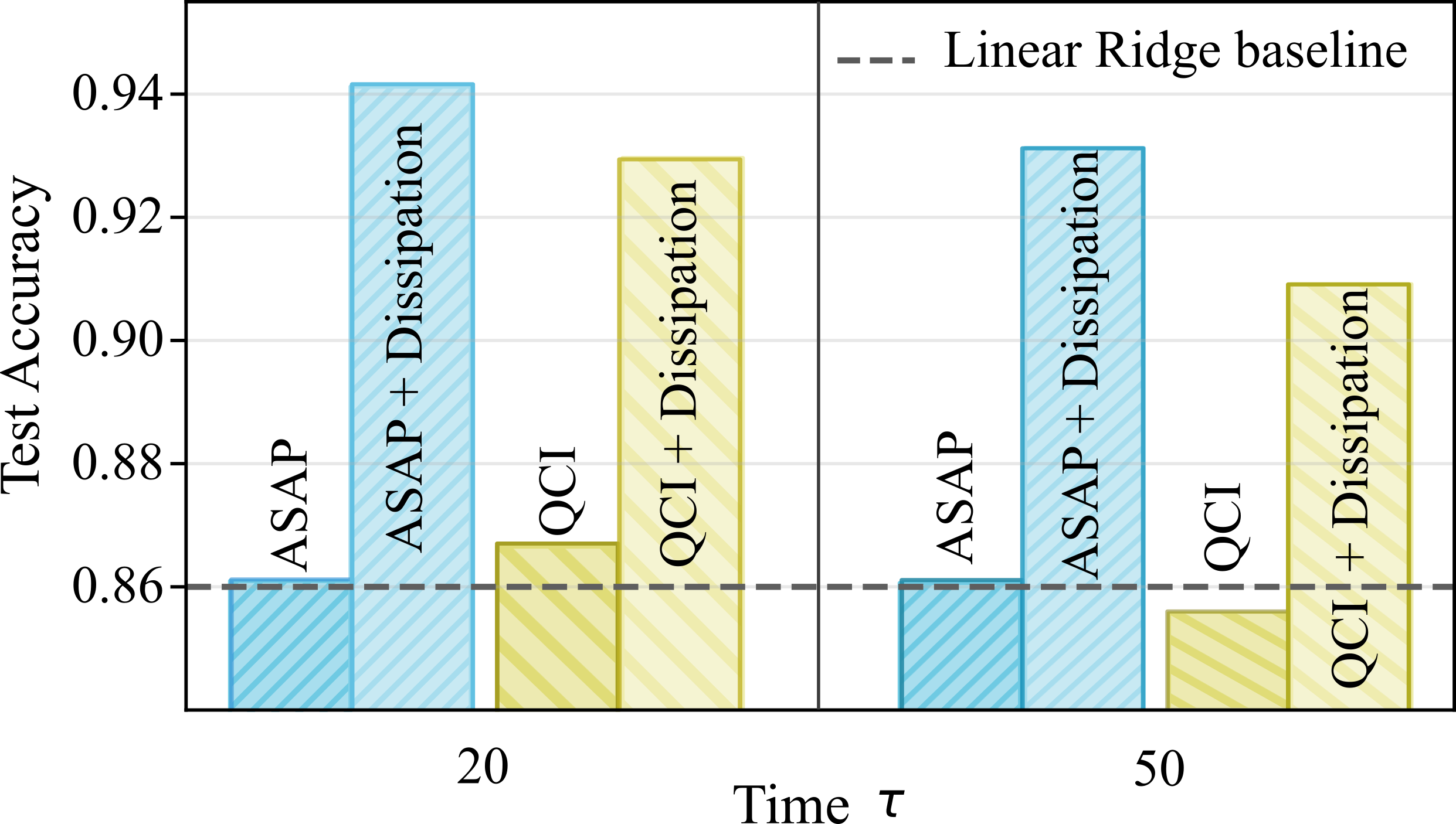}
     \caption{\textbf{Dissipation-induced mitigation of instability driven by many-body quantum scrambling.} Test accuracy of the ASAP and the QCI with and without dissipation at extended evolution times ($\tau = 20$ and $\tau = 50$). The dashed line denotes the linear ridge regression baseline. Absent dissipation, pure unitary evolution subjects the system to many-body quantum scrambling, degrading the test accuracy down to the linear baseline. Conversely, the introduction of controlled dissipation suppresses these instabilities and dramatically recovers learning performance. The dissipation rates are set to $\gamma = 10^{-3}$ for the ASAP and $\gamma = 10^{-2}$ for the QCI. All times are expressed in units of $10\,\text{ns}$.}
     \label{fig:mn23scramble}
 \end{figure}

\textit{Large $\tau$ regime:--} In the large $\tau$ regime, dissipation-free unitary evolution exhibits clear signatures of information scrambling, which leads to a significant degradation in the classification performance of both the ASAP and QCI. This behavior is consistent with the results in Fig.~\ref{fig:mn1all}(b), where the test accuracy begins to decline for $\tau \gg 1$. In Fig.~\ref{fig:mn23scramble}, we perform a comparative analysis between unitary and dissipative dynamics at these extended timescales. Our results demonstrate that the introduction of controlled dissipation significantly restores model performance by suppressing the deleterious effects of quantum scrambling. This suggests that by optimizing the relaxation rate $\gamma$, the HEF can maintain high representational fidelity even in the long-time limit. These findings verify that environmental coupling acts as a stabilizing mechanism, effectively circumventing the "scrambling curse" that typically hinders long-duration reservoir dynamics.

\section{Discussion}\label{sec:discussion}

The results across the three benchmarking stages demonstrate that the Hamiltonian Encoding Framework (HEF) maintains robust and consistent performance across theoretical mapping, analog superconducting evolution (ASAP), and digital gate-based implementations (QCI). Critically, we observe no performance degradation in the analog ASAP realization relative to the digital QCI. From a practical implementation perspective, the ASAP is executed as a single continuous unitary operation, in contrast to the QCI, which requires decomposition into a sequence of discrete quantum gates. Consequently, the ASAP architecture is inherently hardware-efficient and avoids the temporal overhead associated with gate-based composition, offering a promising route toward achieving learning utility on current-generation quantum hardware. Simultaneously, the QCI remains an equally effective implementation strategy, utilizing shallow quantum circuits with an effective depth that scales only linearly with the dimensionality of the input data.

Another major finding of this study is that high-fidelity classification is attainable using only a minimal number of qubits. In comparison, classical neural or reservoir-based models generally require significantly larger architectures—characterized by hidden layers or high neuron counts—to achieve similar performance benchmarks~\cite{zhu2015celm,schaetti2016mnist,pang2016deep}. These results underscore the capacity of the HEF for efficient feature generation, enabling complex data processing even under strict physical resource constraints.

Two distinct feature extraction strategies were evaluated: one based on full density matrix reconstruction and the other on basis measurements. While density matrix features consistently yield superior classification accuracy, the performance differential between the two methods typically remains below $1\%$ across all tested configurations. The marginal advantage of the full density matrix is attributed to its informational completeness, as it captures both population and off-diagonal correlations. In high-accuracy regimes, this advantage becomes negligible. Furthermore, we find that the performance gap can be systematically reduced by incorporating multiple temporal processes or by marginally increasing the qubit count. 

However, this theoretical superiority of the density matrix method must be weighed against the significant measurement overhead. Basis measurements are practically efficient and passive, requiring only standard projective measurements in the computational basis. Conversely, reconstructing the full density matrix needs a rigorous tomographic procedure, requiring active pre-measurement rotations to map coherence information into the measurement basis. This tomographic approach substantially increases the sampling cost and heightens susceptibility to cumulative gate errors and readout noise.
%-------------

Recent investigations have identified a significant scalability challenge in large-scale quantum extreme learning, characterized by exponential concentration phenomena~\cite{Xiong2025, Xiong2025Role}. Consistent with these observations, we find that the proposed HEF experiences performance degradation in the long-time evolution regime due to strong quantum scrambling. This presents a practical hurdle for scaling, as higher-dimensional data requires extended evolution times, which in turn intensifies scrambling effects. However, unlike strict extreme learning architectures, the HEF utilizes the intrinsic dynamical nature of the quantum reservoir. This flexibility permits the introduction of a finite degree of dissipation, which effectively suppresses strong quantum scrambling—a phenomenon substantiated by the results presented in Fig.~\ref{fig:mn23scramble}.

While this study provides a comprehensive analysis, several interesting avenues for future research remain. An area of interest involves the explicit estimation of the cost associated with feature extraction. Additionally, the development of more hardware-efficient training strategies could be useful. For instance, recent advancements in integrated photonics have demonstrated that the readout layer of a quantum extreme learning machine can be trained using high-intensity classical signals and subsequently transferred to quantum-state inference by exploiting the correspondence between stimulated and spontaneous processes~\cite{brusaschi2026quantum}. Although these results were demonstrated on a photonic platform, they suggest that analogous classically assisted or hybrid training protocols could significantly reduce the operational overhead of quantum learning on superconducting systems.

%%%%%%%%%%%%%%%%%%%%%%%%%
\section{Methods}
%%%%%%%%%%%%%%%%%%%%%%%%%
Here, we detail the methodology used to select the effective coupling parameters in the superconducting Hamiltonian and present a principal component analysis of the untrained HEF feature space.

\subsection{Details of Effective Coupling Parameters}\label{appendix1}
The effective exchange interaction between qubits $i$ and $j$ in the reservoir is modeled as
\begin{equation}
J_{ij} = \frac{g_i g_j}{\Delta_{ij}} + \delta J_{ij},
\end{equation}
where $g_i$ represents the coupling strength between qubit $i$ and a shared resonator or coupler mode, and $\Delta_{ij}$ denotes the effective detuning governing the second-order virtual exchange process. 
To capture realistic hardware variability, the coupling strengths $g_i$ are independently drawn from a normal distribution and subsequently lower-bounded in magnitude. This ensures that all qubits maintain a non-trivial connection to the shared interaction channel. Specifically, they are generated according to
\begin{equation}
g_i \sim \mathcal{N}(0.18,\,0.04^2), \qquad g_i \leftarrow \max\bigl(|g_i|,\,0.05\bigr).
\end{equation}
The effective detuning matrix elements $\Delta_{ij}$ for $i \neq j$ are constructed by symmetrizing a matrix of normally distributed variables and applying a constant offset to prevent unphysical resonance singularities:
\begin{equation}
\Delta_{ij} = \left| \frac{\widetilde{\Delta}_{ij} + \widetilde{\Delta}_{ji}}{2} \right| + 0.4, \qquad \widetilde{\Delta}_{ij} \sim \mathcal{N}(1.5,\,0.25^2),
\end{equation}
with the diagonal elements conventionally set to $\Delta_{ii} = 1$. Finally, the correction term $\delta J_{ij}$ accounts for weak, residual direct inter-qubit couplings and structural hardware imperfections. It is introduced as a symmetric perturbation matrix with zero diagonal entries ($\delta J_{ii} = 0$) and small, random off-diagonal elements.

\subsection{Principal Component Analysis of the HEF Feature Space}\label{sec:appendix2}

To systematically quantify the expressivity and effective dimensionality of the Hamiltonian Encoding Framework (HEF) feature space, we perform principal component analysis (PCA) on the extracted dense representations. 

Let $N_t = 60000$ denote the number of training samples and $D = \dim(\mathbf{x}_{\mathrm{dense}})$ be the dimensionality of the feature space. We construct the feature matrix $F \in \mathbb{R}^{N_t \times D}$, where each row corresponds to the HEF feature vector of a single data instance. We first center the feature matrix \cite{greenacre_pca} according to
\begin{equation}
\widetilde{F} = F - \mathbf{1}\mu^{\top},
\end{equation}
where $\mu \in \mathbb{R}^{D}$ is the empirical mean vector of the feature space, and $\mathbf{1} \in \mathbb{R}^{N_t}$ is the all-ones column vector. 

PCA is then executed via the singular value decomposition (SVD) of the centered matrix:
\begin{equation}
\widetilde{F} = U\Sigma V^{\top}.
\end{equation}
Here, $\Sigma$ is a rectangular diagonal matrix containing the singular values ordered in descending magnitude ($\sigma_1 \ge \sigma_2 \ge \ldots \ge 0$). The normalized explained-variance ratio for the $i$-th principal component is given by
\begin{equation}
r_i = \frac{\sigma_i^2}{\sum_{j=1}^{r} \sigma_j^2},
\end{equation}
where $r = \mathrm{rank}(\widetilde{F})$ represents the number of strictly positive singular values. 

Consequently, the cumulative explained variance of the first $k$ principal components is defined as
\begin{equation}
C_k = \sum_{i=1}^{k} r_i.
\end{equation}
The metric $C_k$ concisely quantifies the fraction of the total dataset variance captured within the subspace spanned by the leading $k$ orthogonal principal directions, providing a direct measure of the representational density and expressivity of the HEF.
 
Figure~\ref{fig:mn1}(b) illustrates the cumulative explained variance spectrum obtained from the PCA of the HEF features. From this distribution, we determine the minimum number of principal components required to achieve standard cumulative variance thresholds. Specifically, at a short evolution time of $\tau = 0.2$, capturing 90\%, 95\%, and 99\% of the total variance requires 82, 134, and $290 \pm 20$ principal components, respectively. As the evolution time extends to $\tau = 50$, these requisite dimensionalities increase to 94, 163, and $356 \pm 20$ components.

These results demonstrate that the HEF representation does not suffer from dimensional collapse into a restrictive, low-dimensional subspace; rather, the total feature variation remains distributed across a substantial number of principal directions. This confirms that the high-dimensional feature space generated by the HEF represents a meaningful, expressive manifold capable of rich feature extraction, rather than a trivial low-rank embedding or an informationally redundant over-parameterization.

To systematically evaluate the nonlinear characteristics of the HEF, we analyze both the local curvature of the representation and the discriminative topology of the resulting feature space. For a given input vector $\mathbf{x}$, let $\mathbf{f}(\mathbf{x}) \in \mathbb{R}^{D}$ denote the HEF feature vector extracted following Hamiltonian evolution and readout. We quantify the local nonlinearity by computing a second-order central finite-difference approximation, which acts as a proxy for the norm of the diagonal Hessian elements. For a small perturbation applied along the $j$-th input coordinate, this local response is defined as~\cite{fornberg_fd}:
\begin{equation}
\mathcal{N}_j(\mathbf{x}) = \frac{\left\| \mathbf{f}(\mathbf{x}+\epsilon \mathbf{e}_j) + \mathbf{f}(\mathbf{x}-\epsilon \mathbf{e}_j) - 2\mathbf{f}(\mathbf{x}) \right\|_2}{\epsilon^2},
\end{equation}
where $\mathbf{e}_j$ represents the unit vector along the $j$-th input dimension, and $\epsilon$ denotes the perturbation amplitude. To characterize each data sample, the overall local nonlinearity strength is determined by averaging $\mathcal{N}_j(\mathbf{x})$ over a selected subset of input coordinates $\mathcal{J}$:
\begin{equation}
\mathcal{N}(\mathbf{x}) = \frac{1}{|\mathcal{J}|} \sum_{j\in\mathcal{J}} \mathcal{N}_j(\mathbf{x}).
\end{equation}

Given an ensemble of $M$ sampled inputs $\{\mathbf{x}_m\}_{m=1}^{M}$, the empirical mean local nonlinearity is computed as:
\begin{equation}
\overline{\mathcal{N}} = \frac{1}{M} \sum_{m=1}^{M} \mathcal{N}(\mathbf{x}_m).
\end{equation}
The corresponding standard deviation across the sampled input distribution is given by:
\begin{equation}
\sigma_{\mathcal{N}} = \sqrt{ \frac{1}{M} \sum_{m=1}^{M} \left( \mathcal{N}(\mathbf{x}_m)-\overline{\mathcal{N}} \right)^2 }.
\end{equation}

Across the evaluated evolution times, $\sigma_{\mathcal{N}}$ remains tightly bounded within the range of $23$--$25$, while the mean response $\overline{\mathcal{N}}$ exhibits the upward trend illustrated in Fig.~\ref{fig:mn1}(c). Crucially, if the mapping $\mathbf{f}(\mathbf{x})$ were locally linear or affine, this second-order directional response would vanish identically. Elevated values of $\overline{\mathcal{N}}$ therefore signify strong local curvature, indicating a highly expressive, non-trivial nonlinear feature transformation.

To characterize the discriminative geometry of the HEF feature space, we evaluate the Fisher ratio, as illustrated in Fig.~\ref{fig:mn1}(d). Let $\mu$ represent the global mean of all feature vectors across the dataset, and let $\mu_c$ denote the empirical mean vector of class $c$, which contains $n_c$ samples. The traces of the between-class scatter matrix ($S_B$) and the within-class scatter matrix ($S_W$) are defined~\cite{bishop2006prml} as:
\begin{equation}
\mathrm{Tr}(S_B) = \sum_c n_c \left\| \mu_c-\mu \right\|_2^2,
\end{equation}
and
\begin{equation}
\mathrm{Tr}(S_W) = \sum_c \sum_{\mathbf{f}_i \in c} \left\| \mathbf{f}_i-\mu_c \right\|_2^2,
\end{equation}
where $\mathbf{f}_i$ denotes the feature vector of the $i$-th sample belonging to class $c$. The Fisher ratio is subsequently defined as the quotient of these traces:
\begin{equation}
\mathcal{F} = \frac{\mathrm{Tr}(S_B)}{\mathrm{Tr}(S_W)}.
\end{equation}
A larger Fisher ratio signifies that the inter-class distances (separation of class centroids) are large relative to the intra-class variance (the spread of samples within each respective class). While a high Fisher ratio indicates straightforward linear separability under conventional paradigms, a suppressed untrained Fisher ratio in highly expressive quantum reservoirs does not inherently imply a lack of discriminative utility. Instead, a lower untrained ratio can serve as an indirect indicator of pronounced state-folding and high structural nonlinearity. In this regime, the complex quantum dynamical evolution maps the input data onto a highly folded, nonlinearly intertwined manifold. This causes the class centroids to geometrically converge in the raw feature space—thereby reducing the untrained Fisher ratio—even though the underlying representation contains rich, high-dimensional features that can be effectively decoded by a downstream linearly trained readout.

% \section{Data Availability}
% The data that support the findings of this study were generated using the methods described in the article and are available from the corresponding author upon reasonable request. The MNIST dataset used for benchmarking was obtained from Ref.~\cite{Lecun1998}.

% \section{Code Availability}
% Computer code used to solve the quantum master equation in this study is available from the authors on reasonable request.

\section{Acknowledgment}
S.G. acknowledges funding supports from the National Natural Science Foundation of China (Grant No. 12274034), the Guangdong Province Foreign Experts Project (Flexible Talent Introduction) Fund (No. 2025A1313010019), Shenzhen Specially Appointed Positions (No. 2025TC0140), and University Development Fund (No. UDF01003913). 

% \section{Author Contribution}
% Y.X. and S.G. conceived the project and wrote the paper. Y.X. performed the calculations. Y.X., C.Y. and S.G. analyzed the data, discussed the results, and agreed with the conclusions. S.G. supervised the project.

% \section{Competing interests}
% The authors declare no competing interests.

\bibliographystyle{unsrt}
\bibliography{main_file}

\end{document}